\documentclass[letterpaper]{article} % DO NOT CHANGE THIS
\usepackage{aaai23}  % DO NOT CHANGE THIS
\usepackage{times}  % DO NOT CHANGE THIS
\usepackage{helvet}  % DO NOT CHANGE THIS
\usepackage{courier}  % DO NOT CHANGE THIS
\usepackage[hyphens]{url}  % DO NOT CHANGE THIS
\usepackage{graphicx} % DO NOT CHANGE THIS
\usepackage{natbib}  % DO NOT CHANGE THIS AND DO NOT ADD ANY OPTIONS TO IT
\usepackage{caption} % DO NOT CHANGE THIS AND DO NOT ADD ANY OPTIONS TO IT
\usepackage{algorithm}

\usepackage{booktabs}
\usepackage{mathtools}
\usepackage{float}

\usepackage{algpseudocode}
\usepackage{listings}
\usepackage{bm}
\usepackage{amsmath,amssymb,amsfonts,dsfont,pifont,bm,mathrsfs,mathtools,nicefrac}
\usepackage{color}

\usepackage{newfloat}
\usepackage{listings}
\DeclareCaptionStyle{ruled}{labelfont=normalfont,labelsep=colon,strut=off} % DO NOT CHANGE THIS
\floatstyle{ruled}
\newfloat{listing}{tb}{lst}{}
\floatname{listing}{Listing}
\title{Taming Diffusion Models for Music-driven Conducting Motion Generation}
\author{
    Zhuoran Zhao$^1$\equalcontrib, Jinbin Bai$^1$\equalcontrib, Delong Chen$^2$, Debang Wang$^1$, Yubo Pan$^1$
}
\affiliations{
    $^1$ Dept. of Computer Science, School of Computing, 
    National University of Singapore  \\
    $^2$ Xiaobing.AI
{\{zhuoran.zhao, jinbin.bai\}@u.nus.edu}
}

\usepackage{bibentry}
\begin{document}

\maketitle

\begin{abstract}

Generating the motion of orchestral conductors from a given piece of symphony music is a challenging task since it requires a model to learn semantic music features and capture the underlying distribution of real conducting motion. Prior works have applied Generative Adversarial Networks (GAN) to this task, but the promising diffusion model, which recently showed its advantages in terms of both training stability and output quality, has not been exploited in this context. This paper presents \texttt{Diffusion-Conductor}, a novel DDIM-based approach for music-driven conducting motion generation, which integrates the diffusion model to a two-stage learning framework. We further propose a random masking strategy to improve the feature robustness, and use a pair of geometric loss functions to impose additional regularizations and increase motion diversity. We also design several novel metrics, including Fr\'echet Gesture Distance (FGD) and Beat Consistency Score (BC) for a more comprehensive evaluation of the generated motion. Experimental results demonstrate the advantages of our model. The code is released at \url{https://github.com/viiika/Diffusion-Conductor}.

\end{abstract}
\section{Introduction}

Human conductors have the remarkable ability to translate their rich comprehension of music contents into sequences of precise yet graceful conducting motion. Advancements in AIGC technologies for human motion~\cite{mourot2022survey} have addressed the generation of various human motions such as speech gestures, dance movements, and instrumental motions over recent years, and researchers are now pivoting toward building AI conductors. Pioneered works of VirtualConductor~\cite{chen2021virtualconductor} and M$^2$S-GAN~\cite{liu2022self} demonstrated the promising possibilities of building such systems. These works leverage Generative Adversarial Network (GAN)~\cite{goodfellow2020generative} to learn the probabilistic distribution of real conducting motion from a large-scale paired music-motion dataset. However, GAN-based models typically suffer from notorious issues such as mode collapse and unstable training, which impede the generation of plausible conducting motions.

Recently, diffusion models~\cite{ho2020denoising, ho2022classifier} have emerged as the new state-of-the-art family of deep generative models. Representative models such as GLIDE~\cite{nichol2021glide}, DALL-E 2~\cite{ramesh2022hierarchical}, Latent Diffusion~\cite{rombach2021highresolution}, ImageGen~\cite{saharia2022photorealistic}, and Stable Diffusion~\cite{rombach2022high}, yields impressive performance on conditional image generation, surpassing those GAN-based methods which dominated the field for the past few years. We hypothesize that such an advantage can be extended to the task of music-driven conducting motion generation, and in this paper, we introduce our \texttt{Diffusion-Conductor} model, which is the first diffusion-based AI conductor model.

Our learning framework comprises two consecutive stages, namely the contrastive learning stage and the generative learning stage. The first stage builds a two-tower structure and performs music-motion contrastive pre-training to learn rich music features, those learned features are subsequently transferred to the second stage with a random masking strategy. We incorporate a DDIM-based model to learn the conditional generation of conducting motion, and we modify the supervision signal from $\epsilon$ to $x_0$ for better generation performance. Furthermore, we incorporate perceptual loss to avoid over-smoothing problem and impose additional supervision on the model via two geometric regularization losses, namely velocity loss and elbow loss, to enhance the consistency and diversity of generated motions.

We use a broad array of metrics, including Mean Squared Error (MSE), Fr\'echet Gesture Distance (FGD), Beat Consistency Score (BC), and Diversity, to evaluate the motion produced by \texttt{Diffusion-Conductor}. Thorough comparisons demonstrated that our model outperforms the previous GAN-based method~\cite {liu2022self}. 

In summary, our main contributions are as follows:
\begin{itemize}
    \item Our method is the first work to use diffusion model for music-driven conducting motion generation.
    \item We modify the supervision signal from $\epsilon$ to $x_0$ to achieve the better performance on generating conducting motions, which will inspire later research.
    \item Extensive experiments demonstrate the superiority of our method with quantitative comparison like FGD, BC, and Diversity.
\end{itemize}

\section{Related Works}

\subsection{Audio To Motion Generation}
Recent studies on audio to motion generation can be divided into two categories: speech gesture generation \cite{ahuja2020no, yoon2019robots, liu2022learning, qian2021speech, yoon2020speech} and musical motion generation \cite{liu2022self, yalta2016sequential, li2018skeleton, ren2019music, lee2018listen, sun2020deepdance}. Music-driven conducting motion generation is similar to both of them.
\cite{li2018skeleton, yalta2016sequential} used CNN to extract features from raw music input and fed the extracted features to LSTM network to generate proper body movements. \cite{yoon2019robots} designed an RNN-based encoder and decoder network to generate gesture for a given speech text. To further improve the model's capability, more and more researchers use GAN to generate better results. GAN utilizes adversarial training, where a generator network learns to generate realistic samples by competing against a discriminator network that aims to distinguish between real and generated samples. VirtualConductor \cite{liu2022self} proposed to use GAN to generate conducting motion with sync loss to avoid over-smoothing problem. DeepDance \cite{sun2020deepdance} applied GAN to dance movement generation with additional motion consistency constraints. \cite{liu2022learning, qian2021speech, yoon2020speech} relied on GAN to synthesize speech gesture with adversarial mechanism. However, GAN-based methods often suffer from serious mode collapse and unstable training, which restrict the diversity and quality of generated motion.

\subsection{Diffusion Model}
Diffusion models \cite{ho2020denoising, song2020denoising, dhariwal2021diffusion, ho2022classifier} have emerged as state-of-the-art deep generative models. In this context, a sample from the data distribution is progressively noised in the diffusion process. Subsequently, a deep learning model learns to reverse this process by iteratively denoising the sample. Algorithm~\ref{alg:training} and Alogrithm~\ref{alg:sampling} reveal how to train and inference with denoising diffusion probabilistic models \cite{ho2020denoising}. Diffusion models have demonstrated their potential in various domains, including computer vision, natural language processing, and acoustic signal processing.  Popular examples include GLIDE~\cite{nichol2021glide} and DALL-E 2~\cite{ramesh2022hierarchical} by OpenAI, Latent Diffusion~\cite{rombach2021highresolution} by the University of Heidelberg, ImageGen~\cite{saharia2022photorealistic} by Google Brain and Stable Diffusion~\cite{rombach2022high} by Stability AI. 

Diffusion models have been widely explored in computer vision applications, but there is still limited work on diffusion models in motion applications. DiffGesture \cite{zhu2023taming} has applied diffusion model to audio-driven speech gesture generation. MotionDiffuse \cite{zhang2022motiondiffuse} can be conditioned on text descriptions to generate motions by using diffusion model. To the best of our knowledge, we are the first to use diffusion model for music-driven conductor motion generation. Unlike other motion generation tasks, music-driven conductor motion generation is more complex since conducting motion not only conveys beat information but also expresses articulatory information, such as legato, staccato, etc. We believe that this approach could be a promising direction for future research in the application of diffusion models to motions.

\begin{algorithm}[H]
  \caption{\textbf{Training}} \label{alg:training}
  \small
  \begin{algorithmic}[1]
    \Repeat
        \State $\bm{x}_{0} \sim q(\bm{x}_{0})$
        \State $t \sim \operatorname{Uniform}(\{ 1, \ldots, T\})$
        \State $\epsilon \sim \mathcal{N}(\bm{0},\bm{I}) $
        \State Take gradient descent step on
            \[\nabla_{\theta} \left\| \bm{\epsilon} - \bm{\epsilon}_\theta(\sqrt{\bar\alpha_t} \bm{x}_{0} + \sqrt{1-\bar\alpha_t}\bm{\epsilon}, t) \right\|^2\]
            % $\nabla_{\theta} \left\| \bm{\epsilon} - \bm{\epsilon}_\theta(\sqrt{\bar\alpha_t} \bm{x}_{0} + \sqrt{1-\bar\alpha_t}\bm{\epsilon}, t) \right\|^2$
    \Until{converged}
  \end{algorithmic}
\end{algorithm}

\begin{algorithm}[H]
  \caption{\textbf{Sampling}} \label{alg:sampling}
  \small
\begin{algorithmic}[1]
     \State Trained diffusion model $\theta$, $\bm{x}_T \sim \mathcal{N}(\bm{0}, \bm{I})$
    \For{$t=T, \dotsc, 1$}
        \State $\bm{z} \sim \mathcal{N}(\bm{0},\bm{I})$ if $t > 1$, else $\bm{z} = \bm{0}$
        \State $\bm{x_{t-1}} = \frac{1}{\sqrt{\alpha_t}}\left( \bm{x_t} - \frac{1-\alpha_t}{\sqrt{1-\bar\alpha_t}} \hat{\bm{\epsilon}}_\theta \right) + \sigma_t \bm{z}$
    \EndFor
    \State \textbf{return} $\bm{x}_0$
  \end{algorithmic}
\end{algorithm}

\section{Methods}

In this section, we will explain our task definition, provide an overview of our approach, and illustrate the Training Objective applied at the various stages.

\begin{figure*}[!htbp]
\centering
\includegraphics[width=1\textwidth]{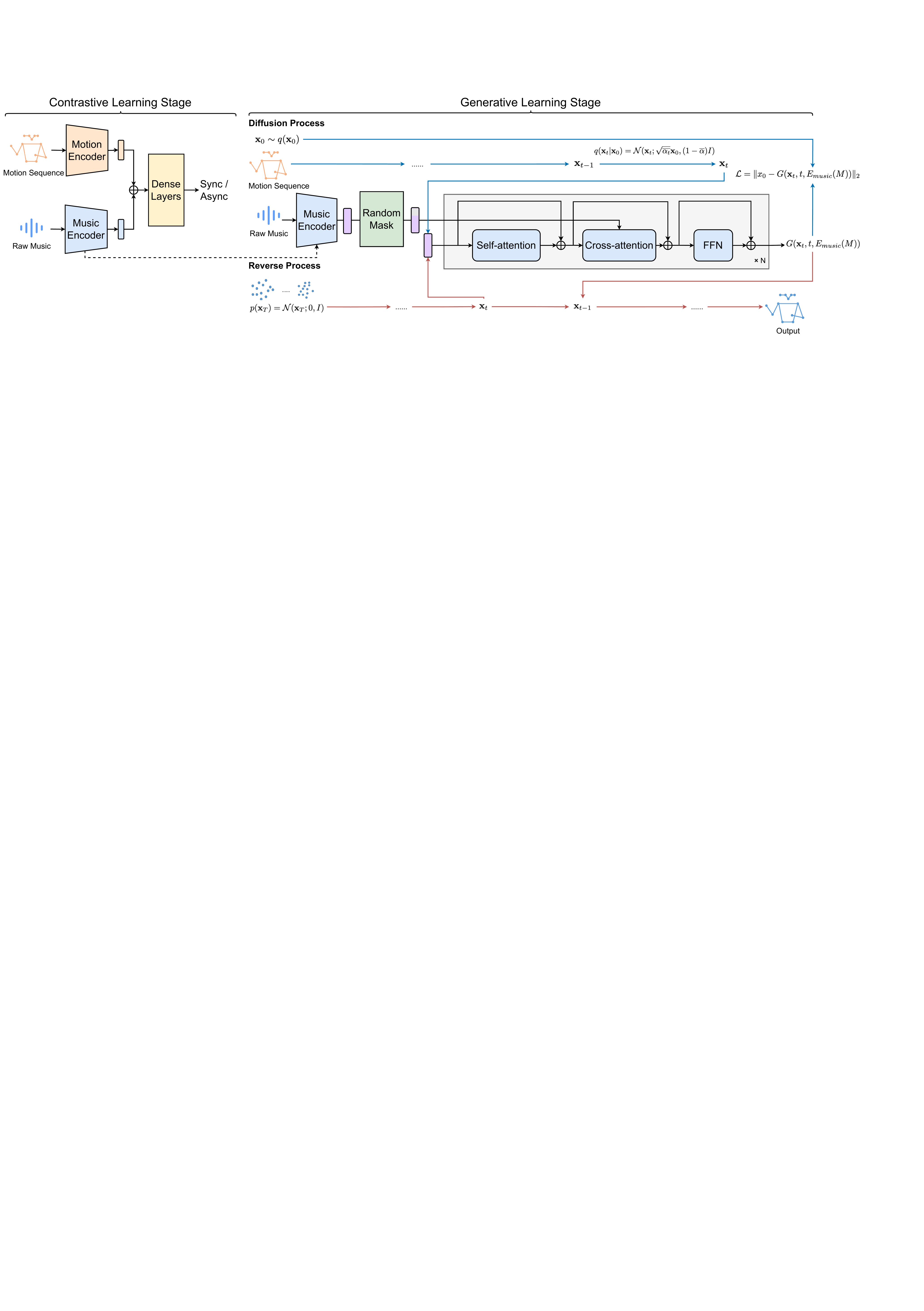}
\caption{\label{fig:architecture}Overview of the proposed framework. The colors of the arrows in Generative Learning Stage represent different stages: blue for training, red for inference, and \textcolor{black}{black} for both training and inference.}
\end{figure*}
\subsection{Task Definition}

Conditional motion generation is a series of tasks that generates realistic and plausible human body motions $x^{1:N}$ with specified actions in response to a given prompt, where $N$ is the length of sequences. The structure of $x^{1:N}$ comprises an array of poses $[x^i]$, where each element $x^i \in \mathbb{R}^{J\times D}$ denotes the pose state at the $i$-th frame, with $J$ being the number of joints and $D$ being the joint dimension. In the case of music-conditioned motion generation, the specified prompt is the music feature $E_{music}(M)$ extracted from the raw music $M$. Our objective is to learn a diffusion model $G$, which can generate a motion sequence $x^{1:N}$ corresponding to the given $E_{music}(M)$.

\subsection{Overview of Our Approach}

Our proposed architecture is illustrated in Fig. \ref{fig:architecture}. In the contrastive learning stage, a contrastive pre-training network composed of a motion encoder $E_{motion}$ and a music encoder $E_{music}$ is used to learn music representations that are correctly aligned to their corresponding motion representations. Subsequently, a generation network $G$ is employed during the generative learning stage to generate a motion sequence based on the music embeddings outputted by the pre-trained $E_{music}$. To further facilitate motion generation while undergoing the denoising process, we make use of the denoising diffusion implicit model (DDIM)~\cite{song2020denoising} and introduce a Cross-Modality Linear Transformer. During inference, a Gaussian distribution noise is sampled according to the given random seed and fed into the denoising process with cross-attention between the music features. Finally, music-driven conducting motions will be generated. Detailed descriptions of our methods are presented in the following sections.

\subsubsection{Contrastive Pre-training}

The contrastive pre-training network comprises three components: a motion encoder $E_{motion}$, a music encoder $E_{music}$, and a set of dense layers $f$. The motion embeddings and music embeddings generated by $E_{motion}$ and $E_{music}$ are concatenated and then passed to $f$, after which a binary cross-entropy loss is applied to assess whether music and motion are appropriately paired. Specifically, the music encoder $E_{music}$ is used to generate music features from raw music and consists of three groups of layers, with each layer comprised of three residual layers and a max-pooling layer. Meanwhile, the motion encoder $E_{motion}$ is employed to generate motion features for the conducting motion sequence. To analyze the conducting motion both spatially and temporally, we make use of the Spatial-Temporal Graph Convolutional Network (ST-GCN) \cite{yan2018spatial}, which has been used extensively in human pose estimation tasks.
\subsubsection{Diffusion Model for Motion Generation}

Diffusion models involve a diffusion process and a reverse process. The diffusion process adds Gaussian noise to the motion sequence data in accordance with the Markov chain rule to approximate the posterior $q(\mathbf{x}_{1:T}|\mathbf{x}_0)$. Upon completion of the diffusion process, the data distribution $\mathbf{x}_T$ should be equivalent to an isotropic Gaussian distribution:
\begin{equation}
    q(\mathbf{x}_{1:T}|\mathbf{x}_0) = \prod\limits_{t-1}^T q(\mathbf{x}_t|\mathbf{x}_{t-1}) 
\end{equation}
\begin{equation}
    q(\mathbf{x}_t|\mathbf{x}_{t-1}) = \mathcal{N}(\mathbf{x}_t; \sqrt{1-\beta_t}\mathbf{x}_{t-1}, \beta_tI)
\end{equation}

Using reparameterization trick, we can sample $\mathbf{x}_t$ at any arbitrary time step $t$ in a closed form:
\begin{equation}
    q(\mathbf{x}_t|\mathbf{x}_0) = \sqrt{\overline{\alpha_t}}\mathbf{x}_0 + \epsilon\sqrt{1-\overline{\alpha_t}}, \epsilon \sim \mathcal{N}(0, I)
\end{equation}

In order to run the reverse process, we need to learn a model $p_{\theta}$ to approximate $q(\mathbf{x}_{t-1}|\mathbf{x}_t)$ since $q(\mathbf{x}_{t-1}|\mathbf{x}_t)$ is intractable:

\begin{equation}
    p_{\theta}(\mathbf{x}_{0:T}) = p(\mathbf{x}_T)\prod\limits_{t-1}^T p_{\theta}(\mathbf{x}_{t-1}|\mathbf{x}_t)
\end{equation}
\begin{equation}
    p_{\theta}(\mathbf{x}_{t-1} | \mathbf{x}_t) = \mathcal{N}(\mathbf{x}_{t-1};\mu_{\theta}(\mathbf{x}_t, t), \Sigma_{\theta}(\mathbf{x}_t, t))
\end{equation}

Most prior works \cite{ho2020denoising, nichol2021glide, zhu2023taming} train the model to predict the noise $\epsilon_{\theta}(\mathbf{x}_t, t, E_{music}(M))$ and then calculate the mean square error between $\epsilon$ and $\epsilon_{\theta}(\mathbf{x}_t, t, E_{music}(M))$ to optimize the model. Here, we instead follow \cite{ramesh2022hierarchical, tevet2022human} by directly predicting the motion $x_0$ and using the mean square error on this prediction which yields better generation performance. Subsequently, the reverse process can be employed to denoise the motion sequence step by step and generate a clean motion sequence conditioned on the given music embeddings.

\subsubsection{Cross-Modality Linear Transformer}

To serve as the denoising model, we make use of a Transformer~\cite{NIPS2017_3f5ee243}. We initially utilize a music encoder to extract the music embeddings, the pre-training of which during the contrastive learning stage can facilitate the generation process. Subsequently, a self-attention module is employed to enable motion features from different times to interact with each other. Additionally, a cross-attention module is utilized to fuse the music embeddings and motion sequence together while a feed-forward network is used to generate motion as $G(\mathbf{x}_t, t, E_{music}(M))$.

\subsubsection{Random Mask}

Inspired by masked language modeling and masked image modeling, we incorporated a random mask~\cite{zhong2020random, tevet2022human} block after the music encoder to train the diffusion model with both music-conditional and unconditional elements. This can potentially allow us to trade off between diversity and quality for improved generalization performance.

\label{sec:3.7}
\subsection{Training Objective}
\subsubsection{Contrastive Learning Stage.}
At the contrastive learning stage, we adopt a binary cross-entropy loss to learn the representation of music under the supervision of motion, which can be formulated as:

\resizebox{0.95\columnwidth}{!}{
\begin{minipage}{\columnwidth}
\begin{align}
\mathcal{L}_{bce} = &\sum_{i,j=1}^{N}( c_{ij} log_2 (f[E_{music}(M_i)\oplus E_{motion}(X_j)]) \nonumber \\
&+(1-c_{ij}) log_2 (1-f[E_{music}(M_i)\oplus E_{motion}(X_j)]))
\end{align}
\end{minipage}
}

where $c_{ij}$ is defined by
$$
c_{i j}= 
\begin{cases}
1, & \text {i = j }\\ 
0, & \text {otherwise }
\end{cases}
$$
$M_i$ and $X_j$ represent the $i$-th music data and the $j$-th motion data respectively, where $\oplus$ denotes the feature concatenation operation. Both $E_{music}$ and $E_{motion}$ denote the music and motion encoders respectively and $f$ represents the dense layers.

\subsubsection{Generative Learning Stage.} The overall training loss for generative learning stage consists of three parts including diffusion loss $\mathcal{L}_{ddim}$, perceptual loss $\mathcal{L}_{perc}$ and geometric loss $\mathcal{L}_{geo}$:

\begin{equation}
    \mathcal{L} = \lambda_{ddim}\mathcal{L}_{ddim} + \lambda_{perc}\mathcal{L}_{perc} + \lambda_{geo}\mathcal{L}_{geo}
\end{equation}

where $\lambda_{ddim}$, $\lambda_{perc}$ and $\lambda_{geo}$ are weighting factors for each loss term.

\subsubsection{Diffusion Loss.}We follow \cite{ramesh2022hierarchical, tevet2022human} to directly predict the motion $x_0$ rather than predicting the noise $\epsilon$ as formulated by \cite{ho2020denoising}, for plausible and improved generation performance. The diffusion loss can be demonstrated as follows:

\begin{equation}
    \mathcal{L}_{ddim} = || x_0 - G(\mathbf{x}_t, t, E_{music}(M)) ||_{2}^{2}
\end{equation}

where $x_0$ is the original motion sequence and $G(\mathbf{x}_t, t, E_{music}(M))$ denotes the final step of motion sequence generated by the diffusion model.

\subsubsection{Perceptual Loss.} Moreover, we employ a perceptual loss to minimize distance between the extracted feature from generated motion and ground-truth motion.

\begin{equation}
    \mathcal{L}_{perc} = \lvert E_{motion}(x_0) - E_{motion}(\hat{x_0}) \lvert
\end{equation}

where $E_{motion}$ is the motion encoder pretrained in the contrastive learning stage and $\hat{x_0}$ equals to $G(\mathbf{x}_t, t, E_{music}(M))$.

\subsubsection{Geometric Loss.} A geometric loss is employed to regularize the generative model, enforcing physical properties and preventing artifacts in order to generate natural and coherent motion. This consists of a velocity loss \cite{tevet2022human} and an elbow loss; the former ensures that the velocity of the generated motion coincides with the ground-truth motion and the latter encourages more intensive arm swing for more vivid motion. The geometric loss is demonstrated as follows:

\begin{equation}
    \mathcal{L}_{geo} = \lambda_{vel}\mathcal{L}_{vel} + \lambda_{elbow}\mathcal{L}_{elbow}
\end{equation}
\begin{equation}
    \mathcal{L}_{vel} = \frac{1}{N-1}\sum_{i=1}^{N-1}||({x_0}^{i+1}- {x_0}^{i}) - (\hat{x_0}^{i+1}- \hat{x_0}^{i})||^2_2
\end{equation}
\begin{equation}
    \mathcal{L}_{elbow} = -\frac{1}{N-1}\sum_{i=1}^{N-1}||\hat{x_0}^{i+1}_{elbow}- \hat{x_0}^{i}_{elbow}||^2_2
\end{equation}

where $\lambda_{vel}$ and $\lambda_{elbow}$ are weighting factors for each term.

\section{Experiment}
In this section, we will first present the training datasets and evaluation metrics. Subsequently, we will conduct quantitative and qualitative experiments that are compared to our baseline method, followed by providing some ablation studies intended to demonstrate the efficacy of our method.

\subsection{Datasets}
We leverage the ConductorMotion100 dataset~\cite{chen2021virtualconductor} for training purposes. It consists of a training set, validation set and test set, with respective durations of 90, 5 and 5 hours. Since the motion of the conductor's lower body contains very little useful information and is often occluded or outside of the camera's view, ConductorMotion only preserves 13 2D keypoints of the upper body in the MS COCO format. All motion data is re-sampled to 30 fps, with corresponding music motion encoding at 90 Hz.

\subsection{Evaluation Metrics}
We use four metrics that are commonly utilized in motion generation and relative fields to evaluate our method.

\subsubsection{Mean Squared Error (MSE).}
Mean squared error (MSE) is the most direct way to measure how closely the generated motion corresponds to the ground-truth motion and has been widely used as an evaluation metric in music-to-motion tasks \cite{kao2020temporally, tang2018dance}. The representation of MSE is defined as follows:

$$MSE(X, \hat{X}) = \| X - \hat{X}\|^2_2$$
where $X$ denotes the ground-truth motion and $\hat{X}$ denotes the generated motion.

\subsubsection{Fr\'echet Gesture Distance (FGD).}
FGD is frequently used to measure the distance between synthesized gesture distribution and real data distribution \cite{zhu2023taming}. Since gesture motion and conducting motion are closely related, both being represented as keypoints, we employ FGD to evaluate the distance of the generated conducting motion distribution and the ground-truth conducting motion distribution. FGD is demonstrated as follows:

$$FGD(Y, \hat{Y}) = \|\mu_{gt} - \mu_{gen}\|^2_2 + \mathrm{Tr}(\Sigma_{gt} + \Sigma_{gen} - 2(\Sigma_{gt}\Sigma_{gen})^\frac{1}{2}$$

where $\mu_{gt}$ and $\Sigma_{gt}$ stand for the mean and variance of the latent feature distribution of the ground-truth motion $X$, while $\mu_{gen}$ and $\Sigma_{gen}$ are the mean and variance of the latent feature distribution of the generated motion $\hat{X}$.

\subsubsection{Beat Consistency Score (BC).}
Beat Consistency Score is a metric to evaluate motion-music correlation in terms of the similarity between the motion beats and music beats. We follow \cite{li2021ai} to define motion beats as the local minima of kinetic velocity and use librosa \cite{mcfee2015librosa} to extract music beats. Beat Consistency Score computes the average distance between every music beat and its nearest motion beat:

$$BC = \frac{1}{|\mathcal{B}^{x}|}\sum^{|\mathcal{B}^{x}|}_{i=1} \exp\Big(-\frac{\min_{\forall{t^x_j}\in\mathcal{B}^{x}}{\|t^x_j - t^y_i\|^2_2}}{2\sigma^2}\Big)$$

where $\mathcal{B}^x = \{t_j^x\}$ represent motion beats and $\mathcal{B}^y = \{t_i^y\}$ represent music beats, and $\sigma$ is the parameter to normalize sequences, which is set to 3 empirically.

\subsubsection{Diversity.}
Similar to prior works \cite{zhu2023taming, li2021ai}, we evaluate our model's ability to generate diverse conducting motions given various input music. Like \cite{zhu2023taming}, we choose 500 generated samples randomly and calculate the mean absolute error between the generated latent motion features and the shuffled features.

\subsection{Implementation Details}
For the diffusion model, we set the diffusion steps to 1000 and use Adam \cite{kingma2014adam} for optimization with a learning rate of 2e-4 and batch size of 48. We train the diffusion model over 500 epochs, setting the unconditional rate of random mask to 0.1. For the weighting factors in the training objective, we set $\lambda_{ddim} = 1$, $\lambda_{perc} = 0.000001$, $\lambda_{geo} = 1$, $\lambda_{vel}=0.1$, $\lambda_{elbow}=0.1$. Experiments are conducted on two NVIDIA TESLA V100 GPUs.

\subsection{Main Results}
\begin{table}[htbp]
  \centering
  \resizebox{\columnwidth}{!}{
    \begin{tabular}{cccccc}
      \toprule
      Methods &  MSE $\downarrow$ &FGD $\downarrow$ & BC $\uparrow$ &Diversity $\uparrow$ \\
      \midrule
      M$^2$S-GAN (VirtualConductor) & 0.0054 & 1051.97 & 0.109 & 1012.06 \\
      Diffusion-Conductor & \textbf{0.0042} & \textbf{812.01} & \textbf{0.119} & \textbf{1152.06}  \\
      \bottomrule
    \end{tabular}
  }
\caption{\label{table1} Main results on ConductorMotion100 test set}
\end{table}
As shown in Table \ref{table1}, we report four metrics compared with VirtualCondutor~\cite{chen2021virtualconductor} on ConductorMotion100 test set. It is shown that our mothod outperforms VirtualConductor on all the four metrics. 

\begin{figure}[htbp]
\centering
\includegraphics[width=\columnwidth]{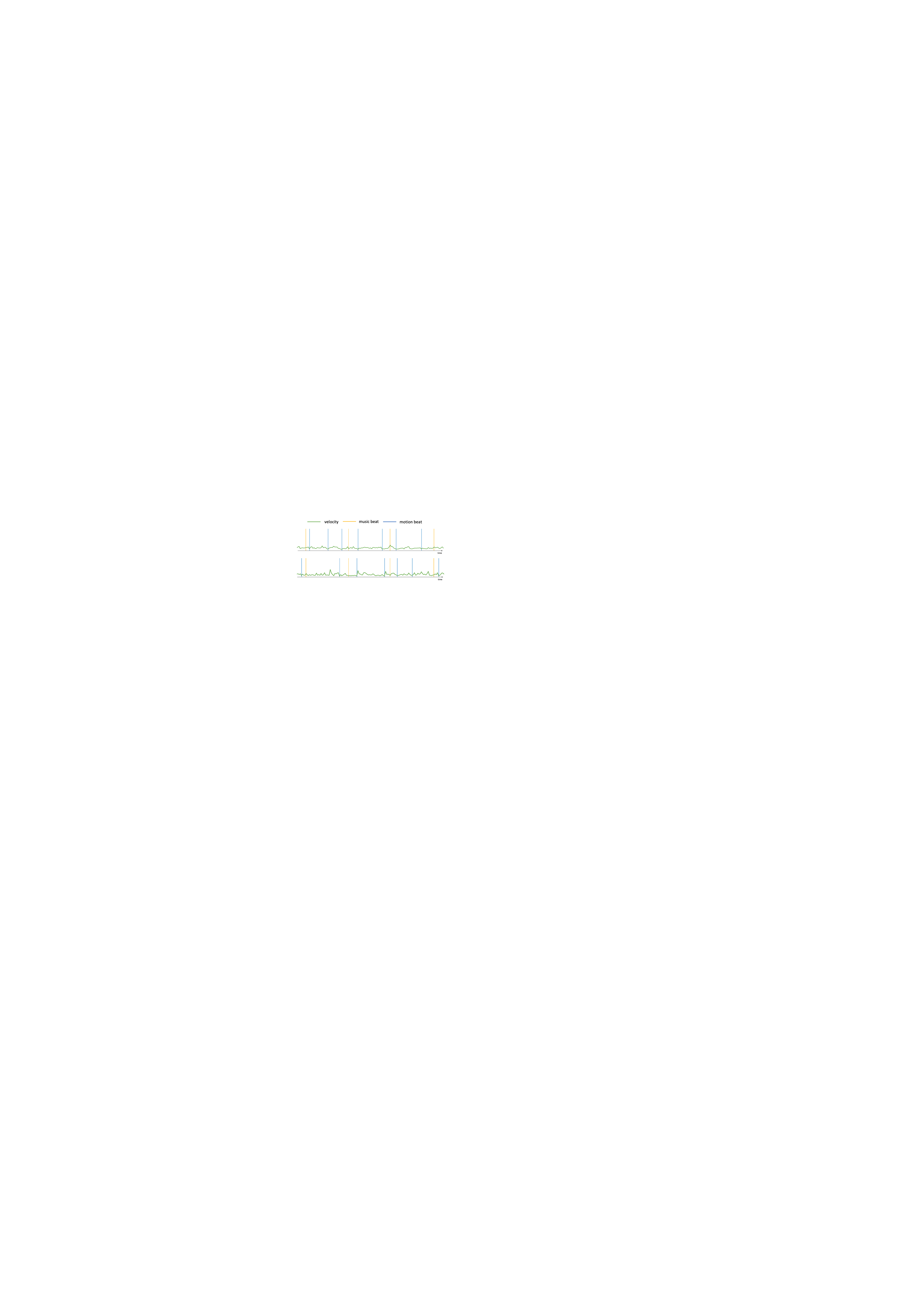}
\caption{\label{fig:beat_consistency} Qualitative comparison of beat consistency between VirtualConductor (top) and ours (bottom).}
\end{figure}

We further visualize the beat consistency between the music and generated conducting motion, making a comparison with VirtualConductor. As illustrated in Fig.\ref{fig:beat_consistency}, our generated motion beats are better able to match the given music beats.

\begin{figure}[htbp]
\centering
\includegraphics[width=0.9\columnwidth]{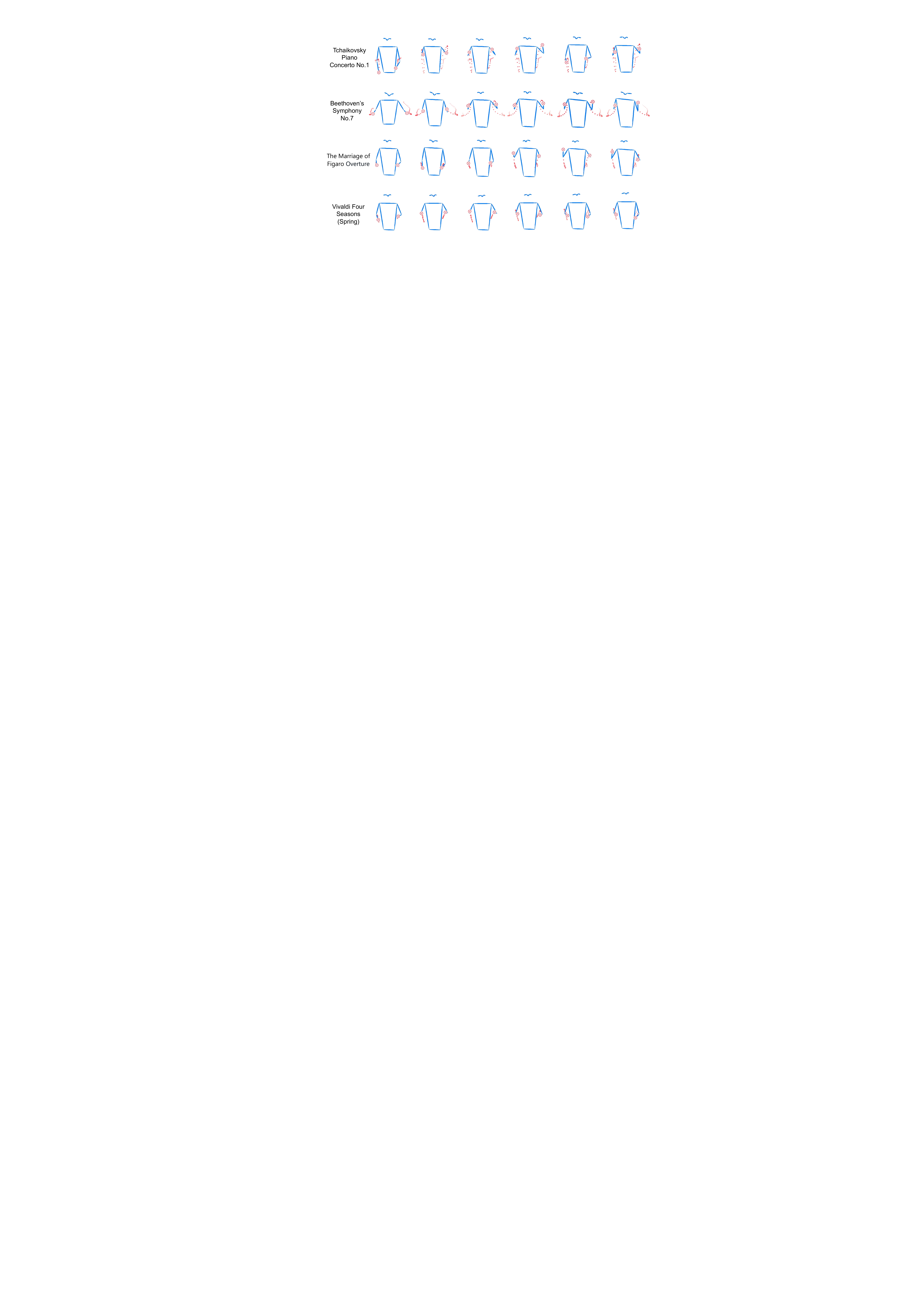}
\caption{\label{fig:visualization} Visualization of the four symphonies.}
\end{figure}

In addition, we provide visualizations of motion generation conditioned on music which were not included in the training or test sets. We randomly select the following symphonies: Tchaikovsky Piano Concerto No.1, Beethoven’s Symphony No.7, The Marriage of Figaro Overture, and Vivaldi Four Seasons (Spring) (see Fig. \ref{fig:visualization}).

\subsection{Ablation Study}
\subsubsection{Comparison of predicting $\epsilon$ and $x_0$.}

\begin{figure}[htbp]
\centering
\includegraphics[width=1\columnwidth]{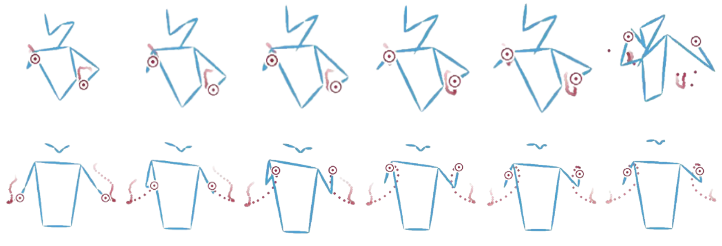}
\caption{\label{fig:visualization_epsilon_x0} Qualitative comparison of generated motion of predicting $\epsilon$ (top) and $x_0$ (bottom) on ConductorMotion100 test set.}
\end{figure}

We further investigate the effect of predicting the noise $\epsilon$ versus the motion $x_0$ via an additional study. The results as indicated in Table. \ref{table2} show that the model trained by minimizing the loss between the noise $\epsilon$ performs much worse than one trained by minimizing the loss between motion $x_0$, which fails to generate plausible motion sequences in longer frames, whereas predicting $x_0$ successfully produces stable and plausible motion sequences (see Fig. \ref{fig:visualization_epsilon_x0}). These results demonstrate the effectiveness of our design-choice to predict the motion rather than noise for each diffusion step.

\begin{table}[htbp]
\centering
  \centering
    \begin{tabular}{ccc}
      \toprule
      Prediction &  MSE $\downarrow$ \\
      \midrule
      $\epsilon$ & 557  \\
      $x_0$ & \textbf{0.0042} \\
      \bottomrule
    \end{tabular}
\caption{\label{table2} Comparison of predicting $\epsilon$ and $x_0$ on ConductorMotion100 test set}
\end{table}

\subsubsection{Effect of geometric loss.}
\begin{figure}[!htbp]
\centering
\includegraphics[width=0.8\columnwidth]{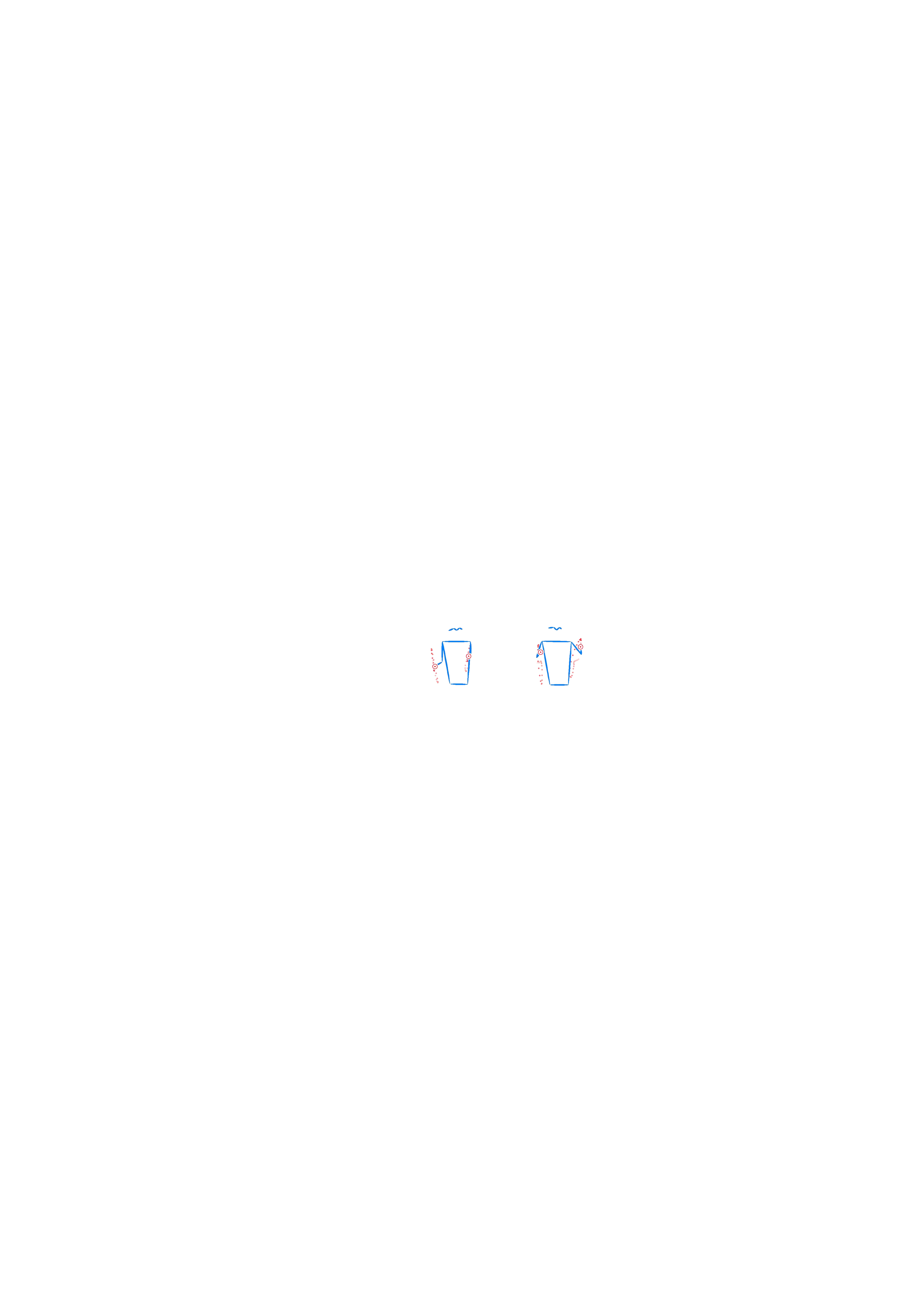}
\caption{\label{fig:geometric_loss} Qualitative comparison of generated motion of w/o (left) and w (right) geometric loss.}
\end{figure}

We examine the effect of incorporating a geometric loss in the training objective and compare it with one trained without its use. The results indicated in Table~\ref{table3} show that the model trained with geometric loss can achieve better performance than the model trained without it on the test set. Furthermore, as visualized in Fig. \ref{fig:geometric_loss}, the model trained with a geometric loss is able to produce motion with more vivid arm swings and plausible poses, which confirms its effectiveness in yielding high-quality motion.

\begin{table}[htbp]
  \centering
  \resizebox{\columnwidth}{!}{
    \begin{tabular}{ccccc}
      \toprule
      Method &  MSE $\downarrow$ &FGD $\downarrow$ & BC $\uparrow$ &Diversity $\uparrow$ \\
      \midrule
      w/o geometric loss & 0.0045  & 822.07 & 0.116 & 1127.90 \\
      w geometric loss & \textbf{0.0042} & \textbf{812.01} &\textbf{0.119} & \textbf{1152.06}  \\
      \bottomrule
    \end{tabular}
  }
\caption{\label{table3} Comparison of five metrics on ConductorMotion100 test set with and without geometric loss}
\end{table}

\section{Conclusion}
In this paper, we presents \texttt{Diffusion-Conductor}, a novel DDIM-based approach for music-driven conducting motion generation, which integrates the diffusion model to a two-stage learning framework. And extensive experiments on several metrics, including Fr\'echet Gesture Distance (FGD) and Beat Consistency Score (BC) demonstrated the superiority of our approach.

\bibliography{aaai23}

\begin{thebibliography}{34}
\providecommand{\natexlab}[1]{#1}

\bibitem[{Ahuja et~al.(2020)Ahuja, Lee, Ishii, and Morency}]{ahuja2020no}
Ahuja, C.; Lee, D.~W.; Ishii, R.; and Morency, L.-P. 2020.
\newblock No gestures left behind: Learning relationships between spoken
  language and freeform gestures.
\newblock In \emph{Findings of the Association for Computational Linguistics:
  EMNLP 2020}, 1884--1895.

\bibitem[{Chen et~al.(2021)Chen, Liu, Li, and Xu}]{chen2021virtualconductor}
Chen, D.; Liu, F.; Li, Z.; and Xu, F. 2021.
\newblock VirtualConductor: Music-driven Conducting Video Generation System.
\newblock \emph{CoRR}, abs/2108.04350.

\bibitem[{Dhariwal and Nichol(2021)}]{dhariwal2021diffusion}
Dhariwal, P.; and Nichol, A. 2021.
\newblock Diffusion models beat gans on image synthesis.
\newblock \emph{Advances in Neural Information Processing Systems}, 34:
  8780--8794.

\bibitem[{Goodfellow et~al.(2020)Goodfellow, Pouget-Abadie, Mirza, Xu,
  Warde-Farley, Ozair, Courville, and Bengio}]{goodfellow2020generative}
Goodfellow, I.; Pouget-Abadie, J.; Mirza, M.; Xu, B.; Warde-Farley, D.; Ozair,
  S.; Courville, A.; and Bengio, Y. 2020.
\newblock Generative adversarial networks.
\newblock \emph{Communications of the ACM}, 63(11): 139--144.

\bibitem[{Ho, Jain, and Abbeel(2020)}]{ho2020denoising}
Ho, J.; Jain, A.; and Abbeel, P. 2020.
\newblock Denoising diffusion probabilistic models.
\newblock \emph{Advances in Neural Information Processing Systems}, 33:
  6840--6851.

\bibitem[{Ho and Salimans(2022)}]{ho2022classifier}
Ho, J.; and Salimans, T. 2022.
\newblock Classifier-free diffusion guidance.
\newblock \emph{arXiv preprint arXiv:2207.12598}.

\bibitem[{Kao and Su(2020)}]{kao2020temporally}
Kao, H.-K.; and Su, L. 2020.
\newblock Temporally guided music-to-body-movement generation.
\newblock In \emph{Proceedings of the 28th ACM International Conference on
  Multimedia}, 147--155.

\bibitem[{Kingma and Ba(2014)}]{kingma2014adam}
Kingma, D.~P.; and Ba, J. 2014.
\newblock Adam: A method for stochastic optimization.
\newblock \emph{arXiv preprint arXiv:1412.6980}.

\bibitem[{Lee, Kim, and Lee(2018)}]{lee2018listen}
Lee, J.; Kim, S.; and Lee, K. 2018.
\newblock Listen to dance: Music-driven choreography generation using
  autoregressive encoder-decoder network.
\newblock \emph{arXiv preprint arXiv:1811.00818}.

\bibitem[{Li, Maezawa, and Duan(2018)}]{li2018skeleton}
Li, B.; Maezawa, A.; and Duan, Z. 2018.
\newblock Skeleton Plays Piano: Online Generation of Pianist Body Movements
  from MIDI Performance.
\newblock In \emph{ISMIR}, 218--224.

\bibitem[{Li et~al.(2021)Li, Yang, Ross, and Kanazawa}]{li2021ai}
Li, R.; Yang, S.; Ross, D.~A.; and Kanazawa, A. 2021.
\newblock Ai choreographer: Music conditioned 3d dance generation with aist++.
\newblock In \emph{Proceedings of the IEEE/CVF International Conference on
  Computer Vision}, 13401--13412.

\bibitem[{Liu et~al.(2022{\natexlab{a}})Liu, Chen, Zhou, Yang, and
  Xu}]{liu2022self}
Liu, F.; Chen, D.-L.; Zhou, R.-Z.; Yang, S.; and Xu, F. 2022{\natexlab{a}}.
\newblock Self-supervised music motion synchronization learning for
  music-driven conducting motion generation.
\newblock \emph{Journal of Computer Science and Technology}, 37(3): 539--558.

\bibitem[{Liu et~al.(2022{\natexlab{b}})Liu, Wu, Zhou, Xu, Qian, Lin, Zhou, Wu,
  Dai, and Zhou}]{liu2022learning}
Liu, X.; Wu, Q.; Zhou, H.; Xu, Y.; Qian, R.; Lin, X.; Zhou, X.; Wu, W.; Dai,
  B.; and Zhou, B. 2022{\natexlab{b}}.
\newblock Learning hierarchical cross-modal association for co-speech gesture
  generation.
\newblock In \emph{Proceedings of the IEEE/CVF Conference on Computer Vision
  and Pattern Recognition}, 10462--10472.

\bibitem[{McFee et~al.(2015)McFee, Raffel, Liang, Ellis, McVicar, Battenberg,
  and Nieto}]{mcfee2015librosa}
McFee, B.; Raffel, C.; Liang, D.; Ellis, D.~P.; McVicar, M.; Battenberg, E.;
  and Nieto, O. 2015.
\newblock librosa: Audio and music signal analysis in python.
\newblock In \emph{Proceedings of the 14th python in science conference},
  volume~8, 18--25.

\bibitem[{Mourot et~al.(2022)Mourot, Hoyet, Le~Clerc, Schnitzler, and
  Hellier}]{mourot2022survey}
Mourot, L.; Hoyet, L.; Le~Clerc, F.; Schnitzler, F.; and Hellier, P. 2022.
\newblock A Survey on Deep Learning for Skeleton-Based Human Animation.
\newblock In \emph{Computer Graphics Forum}, volume~41, 122--157. Wiley Online
  Library.

\bibitem[{Nichol et~al.(2021)Nichol, Dhariwal, Ramesh, Shyam, Mishkin, McGrew,
  Sutskever, and Chen}]{nichol2021glide}
Nichol, A.; Dhariwal, P.; Ramesh, A.; Shyam, P.; Mishkin, P.; McGrew, B.;
  Sutskever, I.; and Chen, M. 2021.
\newblock Glide: Towards photorealistic image generation and editing with
  text-guided diffusion models.
\newblock \emph{arXiv preprint arXiv:2112.10741}.

\bibitem[{Qian et~al.(2021)Qian, Tu, Zhi, Liu, and Gao}]{qian2021speech}
Qian, S.; Tu, Z.; Zhi, Y.; Liu, W.; and Gao, S. 2021.
\newblock Speech drives templates: Co-speech gesture synthesis with learned
  templates.
\newblock In \emph{Proceedings of the IEEE/CVF International Conference on
  Computer Vision}, 11077--11086.

\bibitem[{Ramesh et~al.(2022)Ramesh, Dhariwal, Nichol, Chu, and
  Chen}]{ramesh2022hierarchical}
Ramesh, A.; Dhariwal, P.; Nichol, A.; Chu, C.; and Chen, M. 2022.
\newblock Hierarchical text-conditional image generation with clip latents.
\newblock \emph{arXiv preprint arXiv:2204.06125}.

\bibitem[{Ren et~al.(2019)Ren, Li, Huang, and Chen}]{ren2019music}
Ren, X.; Li, H.; Huang, Z.; and Chen, Q. 2019.
\newblock Music-oriented dance video synthesis with pose perceptual loss.
\newblock \emph{arXiv preprint arXiv:1912.06606}.

\bibitem[{Rombach et~al.(2021)Rombach, Blattmann, Lorenz, Esser, and
  Ommer}]{rombach2021highresolution}
Rombach, R.; Blattmann, A.; Lorenz, D.; Esser, P.; and Ommer, B. 2021.
\newblock High-Resolution Image Synthesis with Latent Diffusion Models.
\newblock arXiv:2112.10752.

\bibitem[{Rombach et~al.(2022)Rombach, Blattmann, Lorenz, Esser, and
  Ommer}]{rombach2022high}
Rombach, R.; Blattmann, A.; Lorenz, D.; Esser, P.; and Ommer, B. 2022.
\newblock High-resolution image synthesis with latent diffusion models.
\newblock In \emph{Proceedings of the IEEE/CVF Conference on Computer Vision
  and Pattern Recognition}, 10684--10695.

\bibitem[{Saharia et~al.(2022)Saharia, Chan, Saxena, Li, Whang, Denton,
  Ghasemipour, Ayan, Mahdavi, Lopes et~al.}]{saharia2022photorealistic}
Saharia, C.; Chan, W.; Saxena, S.; Li, L.; Whang, J.; Denton, E.; Ghasemipour,
  S. K.~S.; Ayan, B.~K.; Mahdavi, S.~S.; Lopes, R.~G.; et~al. 2022.
\newblock Photorealistic text-to-image diffusion models with deep language
  understanding.
\newblock \emph{arXiv preprint arXiv:2205.11487}.

\bibitem[{Song, Meng, and Ermon(2020)}]{song2020denoising}
Song, J.; Meng, C.; and Ermon, S. 2020.
\newblock Denoising diffusion implicit models.
\newblock \emph{arXiv preprint arXiv:2010.02502}.

\bibitem[{Sun et~al.(2020)Sun, Wong, Cheng, Kankanhalli, Geng, and
  Li}]{sun2020deepdance}
Sun, G.; Wong, Y.; Cheng, Z.; Kankanhalli, M.~S.; Geng, W.; and Li, X. 2020.
\newblock DeepDance: music-to-dance motion choreography with adversarial
  learning.
\newblock \emph{IEEE Transactions on Multimedia}, 23: 497--509.

\bibitem[{Tang, Jia, and Mao(2018)}]{tang2018dance}
Tang, T.; Jia, J.; and Mao, H. 2018.
\newblock Dance with melody: An lstm-autoencoder approach to music-oriented
  dance synthesis.
\newblock In \emph{Proceedings of the 26th ACM international conference on
  Multimedia}, 1598--1606.

\bibitem[{Tevet et~al.(2022)Tevet, Raab, Gordon, Shafir, Cohen-Or, and
  Bermano}]{tevet2022human}
Tevet, G.; Raab, S.; Gordon, B.; Shafir, Y.; Cohen-Or, D.; and Bermano, A.~H.
  2022.
\newblock Human motion diffusion model.
\newblock \emph{arXiv preprint arXiv:2209.14916}.

\bibitem[{Vaswani et~al.(2017)Vaswani, Shazeer, Parmar, Uszkoreit, Jones,
  Gomez, Kaiser, and Polosukhin}]{NIPS2017_3f5ee243}
Vaswani, A.; Shazeer, N.; Parmar, N.; Uszkoreit, J.; Jones, L.; Gomez, A.~N.;
  Kaiser, L.~u.; and Polosukhin, I. 2017.
\newblock Attention is All you Need.
\newblock In Guyon, I.; Luxburg, U.~V.; Bengio, S.; Wallach, H.; Fergus, R.;
  Vishwanathan, S.; and Garnett, R., eds., \emph{Advances in Neural Information
  Processing Systems}, volume~30. Curran Associates, Inc.

\bibitem[{Yalta, Ogata, and Nakadai(2016)}]{yalta2016sequential}
Yalta, N.; Ogata, T.; and Nakadai, K. 2016.
\newblock Sequential deep learning for dancing motion generation.
\newblock \emph{Proc. the 46th AI Challenge Study Group}, 43--49.

\bibitem[{Yan, Xiong, and Lin(2018)}]{yan2018spatial}
Yan, S.; Xiong, Y.; and Lin, D. 2018.
\newblock Spatial temporal graph convolutional networks for skeleton-based
  action recognition.
\newblock In \emph{Thirty-second AAAI conference on artificial intelligence}.

\bibitem[{Yoon et~al.(2020)Yoon, Cha, Lee, Jang, Lee, Kim, and
  Lee}]{yoon2020speech}
Yoon, Y.; Cha, B.; Lee, J.-H.; Jang, M.; Lee, J.; Kim, J.; and Lee, G. 2020.
\newblock Speech gesture generation from the trimodal context of text, audio,
  and speaker identity.
\newblock \emph{ACM Transactions on Graphics (TOG)}, 39(6): 1--16.

\bibitem[{Yoon et~al.(2019)Yoon, Ko, Jang, Lee, Kim, and Lee}]{yoon2019robots}
Yoon, Y.; Ko, W.-R.; Jang, M.; Lee, J.; Kim, J.; and Lee, G. 2019.
\newblock Robots learn social skills: End-to-end learning of co-speech gesture
  generation for humanoid robots.
\newblock In \emph{2019 International Conference on Robotics and Automation
  (ICRA)}, 4303--4309. IEEE.

\bibitem[{Zhang et~al.(2022)Zhang, Cai, Pan, Hong, Guo, Yang, and
  Liu}]{zhang2022motiondiffuse}
Zhang, M.; Cai, Z.; Pan, L.; Hong, F.; Guo, X.; Yang, L.; and Liu, Z. 2022.
\newblock MotionDiffuse: Text-Driven Human Motion Generation with Diffusion
  Model.
\newblock \emph{arXiv preprint arXiv:2208.15001}.

\bibitem[{Zhong et~al.(2020)Zhong, Zheng, Kang, Li, and Yang}]{zhong2020random}
Zhong, Z.; Zheng, L.; Kang, G.; Li, S.; and Yang, Y. 2020.
\newblock Random erasing data augmentation.
\newblock In \emph{Proceedings of the AAAI conference on artificial
  intelligence}, volume~34, 13001--13008.

\bibitem[{Zhu et~al.(2023)Zhu, Liu, Liu, Qian, Liu, and Yu}]{zhu2023taming}
Zhu, L.; Liu, X.; Liu, X.; Qian, R.; Liu, Z.; and Yu, L. 2023.
\newblock Taming Diffusion Models for Audio-Driven Co-Speech Gesture
  Generation.
\newblock \emph{arXiv preprint arXiv:2303.09119}.

\end{thebibliography}

\end{document}